\documentclass[aps,prd,twocolumn]{revtex4}
\usepackage{slashed}
\usepackage{amsmath}
\include{./packages}
\usepackage{color}
\usepackage{pstricks}
\usepackage{graphicx}
\usepackage{mathtools}

\begin{document}

\title{  Dilaton in a Multicritical 3+epsilon-D Parity Violating Field Theory }
\author{Gordon W. Semenoff and Riley A. Stewart
\\  Department of Physics and Astronomy, University of British Columbia, 6224 Agricultural Road, 
Vancouver, British Columbia, Canada V6T 1Z1}

\date{}
\begin{abstract}
The multi-critical behaviour of an approximately scale and conformal invariant  quantum field theory, which can be regarded as the deformation of the critical Gross-Neveu model in 3+epsilon dimensions by a nearly marginal parity violating operator, is studied using a large $N$ expansion.  When epsilon is greater than a number of order 1/N, 
the deformation is marginally relevant and it is found to exhibit spontaneous breaking of the approximate scale symmetry accompanied by the appearance of a light scalar in its spectrum. The scalar mass is parametrically small, of order epsilon times the fermion mass and it can be identified with a light dilaton.  When the dimension is reduced to 3 the deformation of the Gross-Neveu model becomes marginally irrelevant, what was a minimum of the potential becomes a maximum and the theory has a non-perturbative global instability.   There is a metastable perturbative phase where the scalar does not condense and the fermions are massless separated by an energy barrier with height of order one (rather than N) from an energetically favoured phase with a runaway condensate. 
\end{abstract}

\maketitle

The possibility that a quantum field theory can have a parameterically small beta function and approximate scale invariance has attracted considerable attention, particularly when the approximate scale symmetry can be spontaneously broken, generating a pseudo-Goldstone boson in the form of a light dilaton. It is one of few mechanisms whereby a scalar field mass is naturally protected from large quantum corrections and it has potentially important phenomenological
applications in the context of the standard model and beyond the standard model building in elementary particle physics  
\cite{Bardeen:1985sm}-\cite{LSD:2023uzj}.  There is also an interesting application of this idea in the physics of cold atoms where it potentially describes an exotic new phase of matter \cite{Semenoff:2017ptn}-\cite{Semenoff:2018yrt}.   

Field theories where such a scenario can be analyzed reliably in a quantitative sense are relatively rare and very recently some new examples 
\cite{Cresswell-Hogg:2022lgg}-\cite{Pomarol:2023xcc} as well as some constraints on finding such examples \cite{Nogradi:2021zqw} have been discussed. In this paper we will study one class of quantum field theory models which has emerged in this recent body of work.  They are quantum field theories containing spinor and scalar fields with scalar self-interactions as well as a Yukawa coupling and they are close relatives of the Gross-Neveu model in spacetimes with dimensions $2<D<4$.  Recent work using the functional renormalization group has suggested that, in the large N limit, such theories can exhibit a line of fixed points along which the coupling constants can be tuned in such a way that the potential energy landscape has a flat direction for a scale symmetry breaking vacuum expectation value of the scalar field \cite{Cresswell-Hogg:2023hdg}.   
What we will find here, using a different technique, agrees with this, as well as some other aspects where we overlap and it extends the analysis to the sub-leading orders of the $1/N$ expansion where the structure is significantly more intricate. We will study the Euclidean quantum field theory with action and Lagrangian density
\begin{align}
&{\bf S}[\phi,\psi,\bar\psi]=\int d^{3+\epsilon}x~\mathcal L (x)\\
&\mathcal L(x)=\sum_{a=1}^Ni\bar\psi_a(x) ( {\slashed\partial}+\phi )\psi_a(x)
 +N  \frac{\lambda_3}{3!}\phi^3 (x)
 \label{L1}
\end{align}
and where expectation values of operators are computed by the Euclidean functional integral
\begin{align}
< ~O~>~=~\frac{ \int [d\phi d\psi d\bar\psi] e^{-{\bf S}[\phi,\psi,\bar\psi]} ~O}{ \int [d\phi d\psi d\bar\psi] e^{-{\bf S}[\phi,\psi,\bar\psi]} }
\end{align}
The spacetime dimension is $D=3+\epsilon$  with $0\leq\epsilon<1$.  The field $\psi(x)$ is a complex Dirac spinor whose number of spinor components times number of flavours is $2N$.      The scalar cubic coupling $ \lambda_3\phi^3(x)$ violates parity which can be defined as the transformation
\begin{align}
\nonumber & {\rm Parity:}~(x_1',x_2',x_3',...)=(-x_1,x_2,x_3,...)\\
&    \psi(x)\to \gamma_1\psi(x') ,~
    \bar\psi(x)\to-\bar\psi(x')\gamma_1,~
    \phi(x)\to -\phi(x')
\label{parity}\end{align}
We will find that, in $3+\epsilon$ dimensions, as long as $0<\epsilon<1$, this model has an analytically accessible, renormalizable $1/N$ expansion.  We will concentrate on the case where $\epsilon$ is small, of order $1/N$, although there is no technical barrier to solving the model for larger values in the interval  $0<\epsilon<1$. 
 It will turn out that, when $\epsilon$ is small and $N$ is large,  this quantum field theory has a relatively flat effective potential as a function of a condensate $<\phi(x)>$ of the scalar field.   This relative flatness is our ``approximate scale invariance''.   In a sense, $\epsilon$ drives the classical violation of the scale invariance that the $D=3$ theory possesses and which the $D=3+\epsilon$  theory does not possess,  and $1/N$ controls the  violation of scale invariance due to quantum mechanical effects. 
We shall find that, depending on the relative magnitudes of $\epsilon$ and $1/N$, $\lambda_3\phi^3(x)$ can either be a marginally relevant or a marginally irrelevant operator.  

As one might anticipate for the deformation of a conformal field theory by a relevant operator,
we shall find that the theory has a stable vacuum with a nonzero condensate of the scalar field in the case
where the deformation$\lambda_3\phi^3(x)$ is marginally relevant.  The nonzero condensate gaps the fermions and, when that occurs, evidenced by a pole in the scalar field propagator, the physical particle spectrum also contains a light scalar field which we identify as a dilaton.  The dilaton mass is parametrically small compared to the fermion mass. This is entirely due to the relative flatness of the potential which we control by keeping $\epsilon$ and $1/N$ small.  

On the other hand, when the operator $\lambda_3\phi^3(x)$ is marginally irrelevant, the model that we are considering has a non-perturbative instability. What was a global minimum of the potential in the above case becomes a global maximum.  There is a metastable local minimum where the condensate is zero and the fermions are massless. It is separated by a potential energy barrier from a more stable phase with a runaway condensate $<\phi>\to\infty$.
The height of the potential barrier between the phases, again due to the relative flatness of the potential, is of order one, (not order $N$ which would help to give the phase a long lifetime).  

In the limit where $\epsilon\to 0$, that is, in exactly three dimensions, the current model has exact scale and conformal invariance at the classical level. 
We will find that it shares many features with another already well-known example of a quantum field theory with broken approximate scale symmetry, the multi-critical three dimensional $O(N)$ model with a $\lambda(\vec\phi^2)^3$ interaction.  The $O(N)$ model has a violation of scale invariance by quantum effects which is weak at large $N$, the beta function being of order $1/N$. We will find that this is so for the theory described by (\ref{L1}) as well. 
 Pisarski \cite{Pisarski:1982vz} showed that the beta function of the $O(N)$ model in the large $N$ expansion has a nontrivial ultraviolet fixed point.  
 It is an example of a quantum field theory which, if $N$ is large enough, is both  infrared free and asymptotically safe. We will find a similar behaviour in the current example. 
 
 At the large $N$ limit, without order $1/N$ corrections taken into account, in both the $O(N)$ model and the one that we consider here, the beta functions vanish and the models (in D=3) are scale invariant for a range of coupling constants.  Moreover, in both cases, there is a certain critical value of the coupling constant where the energy landscape develops a flat direction.  When that occurs, a dimensionful scalar field condensate parameterizes the flat direction and whatever nonzero value the scalar field vacuum expectation value takes up violates scale invariance. This spontaneous breaking of scale invariance is accompanied by a massless dilaton which one can view as simply the fluctuation of the scalar field along the flat direction.  Both the phase transition and the presence of the dilaton were already noted for the $O(N)$ model by Bardeen, Moshe and Bander \cite{BMB}. 
 Finally, in both cases, as is already known for the O(N) model \cite{Omid:2016jve} and as we shall see here for the current model when the dimension is exactly three and once $1/N$ corrections are included, stability is at issue.   For the $O(N)$ model, it is only known that the extremum of the potential with $1/N$ corrections turned on becomes a local maximum. It is not known whether or not there are other stable states.  For the current model, the minimum of the potential is at $<\phi>\to\infty$ which decouples the fermions and the scalar field mass also diverges. 
   
    The scalar field $\phi(x)$ in (\ref{L1}) has no kinetic term and it has an unconventional scaling dimension.  The absence of a kinetic term, which would be an operator with classical dimension four, makes conventional perturbation theory difficult.  However, this theory is ideally suited to be studied in an asymptotic expansion at large $N$ where we expand order by order in $1/N$.  Moreover, the expansion in $1/N$ is thought to be renormalizable.  We will confirm that this is indeed so for the leading orders of the $1/N$ expansion.   We will renormalize the model in equation (\ref{L1}) by adding counterterms which will cancel any occurrence of the operators $i\bar\psi(x)\psi(x)$, $\phi(x)$ or $\phi^2(x)$ in the effective action. We will also renormalize the Yukawa vertex $i\bar\psi(x)\phi(x)\psi(x)$ using a wave-function renormalization of the scalar field $\phi(x)\to Z^{1/2}\phi(x)$. Then, finally, we will renormalize $\lambda_3\phi^3(x)$ using a subtraction scheme which puts the effective potential in a convenient form.  
  
  We note that this field theory, being parity violating, has no symmetry which suppresses a vacuum expectation value of the scalar field $<\phi>$ or the radiative generation of a fermion mass.  
  Our renormalization scheme which cancels the fermion mass operator $\sim i\bar\psi\psi$ and the scalar field tadpole $\sim\phi$ is equivalent to defining $\phi$ as that scalar field whose condensate $<\phi>$ equals the fermion mass.  Or, equivalently, requiring a covariance, rather than invariance, of the effective action under a parity transformation ${\bf S}_{\rm eff}[<\phi>,\lambda_3]= {\bf S}_{\rm eff}[-<\phi>,-\lambda_3]$.  Since there is no symmetry mechanism which protects the operators $\phi$ or $i\bar\psi\psi$, we expect that there is always, eventually a nonzero condensate  $<\phi>\neq 0$.  Indeed, we will find that this is so in every case where this is a consistent quantum field theory.  What is special here is the relative flatness of the energy landscape near the condensate which leads to the parametrically small mass of the scalar field itself. 
   
To implement the large $N$ expansion, we integrate out the fermion field to get the effective action for the scalar field,
\begin{align}
{\bf S}_{\rm eff}[\phi]&= -N{\rm Tr}\left(i\slashed\partial +i\phi\right)+\int d^{3+\epsilon}x  \frac{N\lambda_3}{6}\phi^3(x)
 \label{seff}\end{align}
When $N$ is large, the remaining functional integral, which is over the scalar field, can be analyzed using the saddle-point approximation.  
For this purpose, we write $\phi(x)=\phi_0+\frac{\delta\phi(x)}{\sqrt{N}}$ where $\phi_0=<\phi>$, the condensate of the scalar field.  We shall assume that $\phi_0$ is an $x$-independent constant.  Then we expand
the effective action to second order in $\delta\phi(x)$ and, anticipating that $\phi_0$ will be adjusted to
minimize the action, we drop the linear term in $\delta\phi(x)$.  The result is 
\begin{align}
 {\bf S}_{\rm eff}\left[\phi_0+\frac{\delta\phi}{\sqrt{N}}\right]  
 &=N\int d^{3+\epsilon}x \left(\frac{ \Gamma[-\frac{3}{2}-\frac{\epsilon}{2}]| \phi_0|^{3+\epsilon}}{(4\pi)^{\frac{3}{2}+\frac{\epsilon}{2}}} + \frac{\lambda_3\phi_0^3}{6}\right) 
\nonumber \\ &+
 \frac{1}{2}\int d^{3+\epsilon}xd^{3+\epsilon}y \delta\phi(x)\Delta^{-1}(x,y)\delta\phi(y)
 \nonumber \\ &+
 \ldots \label{seff1}
\end{align}
where the ellipses stand for contribution that are of order $1/N^{\frac{1}{2}}$ and higher. The quadratic term
in $\delta\phi$ in equation (\ref{seff1}) is the inverse scalar field propagator (with a factor of $N$ removed),
\begin{align}
&\Delta^{-1}(x,y)= \lambda_3\phi_0 \delta(x-y)
  -{\rm Tr}(x|\frac{1}{i\slashed\partial+i\phi_0}|y)(y|\frac{1}{i\slashed\partial+i\phi_0}|x)
\label{delta}\\
&\Delta^{-1}(p)=\lambda_3 \phi   - \nonumber \\ &
 \frac{2(2+\epsilon) \Gamma\left[-\frac{1}{2} - \frac{\epsilon}{2}\right](\frac{p^2}{4} + \phi^2)^{\frac{1}{2} + \frac{\epsilon}{2}} 
}{(4 \pi)^{\frac{3}{2} + \frac{\epsilon}{2}}}
 {}_{2}F_{1}\left(\tfrac{1}{2}, -\tfrac{1}{2}-\tfrac{\epsilon}{2}, \tfrac{3}{2}, \tfrac{p^2}{4 \phi^2 + p^2} \right)
\label{delta(p)}\end{align}
where ${}_{2}F_{1}$ is the Gauss hypergeometric function. 

\begin{figure}
\begin{center}
\includegraphics[scale=1.5]{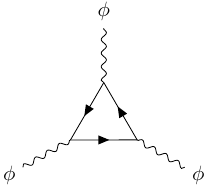}
 \caption{A discrete analog of the parity anomaly \cite{Niemi:1983rq}-\cite{Redlich:1983dv}:~ The triangle diagram has fermion lines in the internal loop, the vertices are Yukawa interactions and the three external lines are scalar fields.  Due to parity symmetry, this graph vanishes in the limit where the parity violating fermion mass is put to zero while keeping the external momenta nonzero. 
  However, it is non-zero when the parity violating mass is non-zero and in that case, it goes to a constant, $\sim |{\rm mass}|^{\epsilon}{\rm sign}$(mass), when the external momenta are put to zero while holding the mass nonzero. This leads to the $\sim N |\phi_0|^{3+\epsilon} $ contribution of the fermion determinant to the effective action. In exactly D=3, when $\epsilon=0$ we see that the order of limits, momentum to zero then mass to zero leaves a  nonzero value for this 3 point function whereas for the opposite order the result would be zero.
  This is so when $\lambda_3=0$ where this model reduces to the critical $D=3$ Gross-Neveu model.  The symmetry breaking phase of the Gross-Neveu model in dimension $D=3$ would not exist without this anomaly contribution.}
\label{fig1}
\end{center}
\end{figure}

The non-analytic structure of the first term  $\sim|\phi_0|^{3+\epsilon}$ of the right-hand-side of equation (\ref{seff1})  is a result of non-analyticity of the fermion determinant in the region near zero fermion mass. (See figure \ref{fig1} for a discussion of an  implication of this non-analyticity.)
  We have used dimensional regularization to compute the fermion determinant and, for constant $\phi_0$, the term $\sim|\phi_0|^{3+\epsilon}$ is the only contribution and it is not ultraviolet divergent.  In dimensional regularization, terms $\sim \phi_0$ or $\sim \phi_0^2$ are absent. If they were present, as they might be with other regularizations, we assume that they are entirely canceled by counter-terms.  

To proceed, it is instructive to first examine the limit where $\epsilon\to 0$, that is,  spacetime dimension   $D=3$, where the coupling constant $\lambda_3$ becomes dimensionless. Let us examine this special case in more detail.  The  first two terms, which are of order $N$, occurring on the right-hand-side  of equation (\ref{seff1}) exhibit exact scale invariance. Let us define the ``$D=3,N\to\infty$'' limit of this theory as what we would obtain by truncating the effective action to the terms that are displayed in equation (\ref{seff1}). Then scalar field takes up that value of $\phi_0$ which minimizes the effective potential
 \begin{align}
 {\bf V}_{\rm eff}^{D=3,N\to\infty}[\phi_0]  
 &=N \left( \frac{1}{6\pi} + \frac{\lambda_3{\rm sign}(\phi_0)}{6}\right) | \phi_0|^{3}
 \label{seff13d}
\end{align}
(We remind the reader that the effective potential is equal to the effective action evaluated on a constant field and divided by the spacetime volume.)
 We see that, for values of $\lambda_3$ in the range $\frac{-1}{\pi}<\lambda_3<\frac{1}{\pi}$, the effective action is minimized at $\phi_0=0$.  
 When $|\lambda_3|> \frac{1}{\pi}$ the effective action is unbounded from below. If we fine tune to $\lambda_3=\frac{1}{\pi}$ or $\lambda_3=-\frac{1}{\pi}$ the effective action is flat -- $\phi_0$-independent -- and the condensate can take up any value of $\phi_0$.  When it takes up a particular value, the exact scale symmetry of this ($D=3,N\to\infty$ model) is spontaneously broken.
 It is also easy to see that the scalar field propagator $\Delta(p)$, in that case, where it has the form
 \begin{align}
  \Delta^{D=3,N\to\infty}(p)=\frac{12\pi|\phi_0|} { p^2\left(1-\frac{1}{20}\frac{p^2}{\phi_0^2}+\frac{12}{2240}\frac{p^4}{\phi_0^4}+\ldots\right) }
   \end{align}
  has a pole at $-p^2=0$ whenever $\phi_0$ is nonzero. This pole is due to the massless dilaton which is a Goldstone boson resulting from the breaking of exact scale symmetry. 
  

Before we explore $1/N$ corrections, let us examine the possibility of staying at infinite $N$ and detuning the scale invariance by going to $3+\epsilon$ dimensions where the coupling $\lambda_3$ has classical dimension $\epsilon$.    The effective potential becomes
 \begin{align}
 {\bf V}_{\rm eff}^{D=3+\epsilon,N\to\infty}[\phi_0]  
 &=N\left(\frac{ \Gamma[-\frac{3}{2}-\frac{\epsilon}{2}]| \phi_0|^{3+\epsilon}}{(4\pi)^{\frac{3}{2}+\frac{\epsilon}{2}}} + \frac{\lambda_3\mu^\epsilon\phi_0^3}{6}\right) 
 \label{seff13d+epsilon}
\end{align}
 where we have replaced $\lambda_3$ by $\lambda_3\mu^\epsilon$ where $\lambda_3$ is now dimensionless and $\mu$ is a dimension one parameter which we will later on identify with a renormalization scale. Note that the coupling constant has a positive scaling dimension in total which implies that it is a relevant operator. 
The effective potential has two terms with different powers of $\phi_0$.  It is always stable -- bounded from below -- when $\epsilon>0$ and for any value of $\lambda_3$. 
Moreover, there is always a minimum at a non-zero value of $\phi_0$, 
\begin{align}
\hat \phi_0~=~-\mu\cdot {\rm sign}(\lambda_3)\left( \frac{(4\pi)^{\frac{3}{2}+\frac{\epsilon}{2}}| \lambda_3| }{-4 \Gamma[-\frac{1}{2}-\frac{\epsilon}{2}]}\right)^{\frac{1}{\epsilon}}
\label{large N solution}
\end{align}
When $\epsilon$ is small, the exponent on the last factor in equation (\ref{large N solution}) is large 
and this equation might be more sensibly written as an equation for that value of $\lambda_3$ which is 
required to achieve a particular value of $\phi_0$, 
\begin{align}
\lambda_3=  -{\rm sign}(\phi_0)\frac{-4 \Gamma[-\frac{1}{2}-\frac{\epsilon}{2}]}{(4\pi)^{\frac{3}{2}+\frac{\epsilon}{2}}  }
\left|\frac{\phi_0}{\mu}\right|^\epsilon
\label{large N solution1}
\end{align}
When $\epsilon$ is very small, 
\begin{align}
\lambda_3\to  -{\rm sign}(\phi_0)\frac{1}{\pi}+\mathcal O(\epsilon)
\end{align}
reproduces the critical coupling that was found in $D=3$. 

The curvature of the effective potential at its minimum is 
\begin{align}
 \left. { {\bf V}_{\rm eff}^{D=3+\epsilon,N\to\infty}}''
 \right|_{{ {\bf V}_{\rm eff}^{D=3+\epsilon,N\to\infty}}'=0}
 =N\epsilon \left(\tfrac{-4 \Gamma[-\frac{1}{2}-\frac{\epsilon}{2}]  }{(4\pi)^{\frac{3}{2}+\frac{\epsilon}{2}} } \right)| \hat\phi_0|^{1+\epsilon}
\end{align}
This curvature is of order $\epsilon N$, rather than  $N$ that  would generically be expected. 
For small $\epsilon$, the pole in $\Delta(p)$ occurs at 
\begin{align}
-p^2\approx 12\epsilon\hat\phi_0^2 +\mathcal O(\epsilon^2,\epsilon/N,1/N^2)
\end{align}
which is the due to the dilaton.  For small $\epsilon$, the dilaton mass $\sqrt{12\epsilon}|\hat\phi_0|$ is parametrically small compared to the fermion mass which is equal to $|\hat\phi_0|$. 

Now that we have studied the large N limit, let us examine the leading corrections in $1/N$.  To  find the radiative correction to the effective action (\ref{seff1}) we must include the result of the gaussian integration over the scalar field $\delta\phi$.  We will do this with the assumption that $\epsilon\sim \frac{1}{N}$, so that, in the next-to-leading order in $1/N$, we can set $\epsilon$ to zero.  What we find here is easy to generalize to larger $\epsilon$.  
The effective potential becomes
 \begin{align}
 {\bf V}_{\rm eff}  =N&  \left(\frac{ \Gamma[-\frac{3}{2}-\frac{\epsilon}{2}]| \phi_0|^{3+\epsilon}}{(4\pi)^{\frac{3}{2}+\frac{\epsilon}{2}}} + \frac{\lambda_3\mu^\epsilon\phi_0^3}{6} \right) \nonumber \\ &
+
   \frac{1}{2}\int  \frac{d^{3}p}{( 2\pi)^{3}} \ln \left[\Delta^{-1}(p, \phi_0)\right] 
   +\mathcal O(1/N)
 \label{seff2}
 \end{align}
 where $\Delta(p,\phi_0)$ is the scalar two-point function at order $1/N$ given in equation (\ref{delta(p)}) with $\epsilon$ set to zero. 
 
 The momentum integral in the correction term is ultraviolet divergent and it must be regularized and  renormalized.
 Accordingly, after dropping a divergent, $\phi_0$-independent term, we use an asymptotic expansion 
 at large $p$ of the remaining integrand, $\ln\left[\frac{8}{p}\Delta^{-1}(p,\phi_0)\right]$,  up to order $1/p^3$. We get
\begin{align}
&\frac{1}{2}\int\frac{d^3p}{(2\pi)^3}\ln\left[ \frac{8}{p}\Delta^{-1}(p,\phi_0)\right]=\frac{1}{4\pi^2}\int_0^\Lambda p^2dp \left(\frac{A}{p}+\frac{B}{p^2}\right) \nonumber \\ &
+
\biggl\{-  \frac{16}{ \pi^2}\frac{|\phi_0|^3}{6\pi}
 +\left[ 
\frac{2^8}{\pi^2} \lambda_3^3   - \frac{48}{\pi^2}  \lambda_3\right]\frac{\phi_0^3}{6}\biggr\} \int_0^\Lambda  \frac{p^2dp}{p^2(p+|\phi_0|)}
\nonumber \\ &+\frac{1}{4\pi^2}\int_0^\Lambda p^2dp \left(\ln\left[ \frac{8}{Np}\Delta^{-1}(p,\phi_0)\right] -(``\ldots")\right)
\label{regularization}
\end{align}
In the first  and second lines of equation (\ref{regularization}) we have taken an asymptotic expansion of the integrand at large $p$ and we displayed the terms which, when integrated, will be ultraviolet divergent. To regulate them, we have imposed a large momentum cutoff, $\Lambda$.   
Note that we have replaced $\frac{1}{p^3}$ by $\frac{1}{p^2(p+|\phi_0|)}$ in order to make the integral of this term converge in the infrared, at $p\sim 0$.  We have chosen $|\phi_0|$ as the infrared cutoff there simply to avoid introducing new dimensionful parameters.   
The integral of the third line in the above equation (\ref{regularization}) has these divergent terms subtracted and it is now finite.  In it, the ultraviolet cutoff can removed and its dependence on the remaining dimensionful parameter, $\phi_0$ is determined by dimensional analysis, producing the terms  
\begin{align}
\frac{1}{4\pi^2}\int_0^\infty p^2dp \left(\ln\left[ \frac{8}{Np}\Delta^{-1}(p,\phi_0)\right] -(``\ldots")\right)
\nonumber \\ =\frac{\zeta_1(\lambda_3)}{6\pi}|\phi_0|^3+\frac{\zeta_2(\lambda_3)}{6}\phi_0^3
\end{align} 
where  $\zeta_1(\lambda_3) $ and $\zeta_2(\lambda_3) $ are finite functions of $\lambda_3$ which we can easily find integral expressions for, but we shall not need their explicit forms.

We assume that the quadratic and linearly divergent integrals can be canceled by local counter-terms which are linear and quadratic in $\phi_0$, respectively (after confirming the nontrivial fact that they are actually analytic in $\phi_0$ itself). The logarithmic divergent part contains a term which is proportional to  $|\phi_0|^3$ which, since we are not allowed a counter-term that is non-analytic in $\phi_0$,  must be canceled by wavefunction renormalization. 
 \begin{align}
 \phi_0 \to Z^{\frac{1}{2}}\phi_0 = \phi_0\left( 1-\frac{\zeta_1(\lambda_3)}{3N}+\frac{16}{3\pi^2N}\ln\frac{\Lambda}{\mu} +\mathcal O(1/N^2)\right)
 \label{Z}\end{align}
 Once we have implemented this wavefuntion renormalization, there remains a divervent part $\sim\phi_0^3$ which is canceled by coupling constant renormalization
  \begin{align}
 \lambda_3\to \lambda_3 - (\zeta_2(\lambda_3)-\zeta_1(\lambda_3) )- \frac{32}{\pi^2N} (
8 \lambda_3^3 -  \lambda_3)\ln\frac{\Lambda}{\mu} 
 \label{rlambda}\end{align}

 These renormalizations are determined by the subtraction scheme where the positive dimensionful constant $\mu$
 is the value of $|\phi_0|$ where the effective potential  is given by its leading order in large $N$,
 \begin{align}
 {\bf V}_{\rm eff}[\phi_0=-\mu\cdot{\rm sign}(\lambda_3)]~=~N  \biggl\{   \frac{ \Gamma[-\frac{3}{2}-\frac{\epsilon}{2}]}{(4\pi)^{\frac{3}{2}+\frac{\epsilon}{2}}} - \frac{|\lambda_3|}{6}   
\biggr\} \mu^{3+\epsilon}
 \label{scheme}
 \end{align}

  The result for the effective potential is then
 \begin{align}
 &{\bf V}_{\rm eff}  =N \biggl\{   \frac{ \Gamma[-\frac{3}{2}-\frac{\epsilon}{2}]|\phi_0|^{3+\epsilon}}{(4\pi)^{\frac{3}{2}+\frac{\epsilon}{2}}} + \frac{\lambda_3\mu^\epsilon\phi_0^3}{6} 
    \biggr\}\nonumber \\ &
+\biggl\{-  \frac{16}{ \pi^2}\frac{|\phi_0|^3}{6\pi}
 +\left[ 
\frac{2^8}{\pi^2} \lambda_3 ^3   - \frac{48}{\pi^2}  \lambda_3\right]\frac{\phi_0^3}{6}\biggr\}\ln\frac{\mu}{|\phi_0|}
\nonumber \\ & ~~~~~~~~~~~~~~~~~~~~~~~ +\mathcal O(1/N)
\label{renormalized V} \end{align}
 The logarithm contains the subtraction scale $\mu$.  
 The appearance of $|\phi_0|$ in the logarithmic term limits the range of validity of perturbation theory.  In fact  $\phi_0\to0$ is no longer in the perturbative regime since the correction term would be large.
  
 The domain of validity of the large $N$ computation can be improved by using the renormalization group which
 resums contributions which contain higher powers of $\frac{1}{N}\ln|\phi_0|/\mu$ in such a way that the effective potential satisfies the renormalization group equation. This follows a procedure outlined by Coleman and Weinberg in their seminal work on mass scale generation in a tree-level scale invariant field theory \cite{Coleman:1973jx}. To this end, we deduce the 
renormalization group functions from equations (\ref{Z}) and (\ref{rlambda}),
 \begin{align}
\gamma=\frac{16}{3\pi^2N} 
 ~~,~~
\beta_3(\lambda_3)=-\epsilon\lambda_3 +\frac{32}{\pi^2N} (\lambda_3-
8 \lambda_3^3) 
\label{beta function}
 \end{align}

To proceed, we demand  that the effective action, with the subtraction scheme defined in equation (\ref{scheme}),  satisfies the renormalization group equation 
 \begin{align}
 \left(\mu\frac{\partial}{\partial\mu}+\gamma \phi_0 \frac{\partial}{\partial\phi_0}
 +\beta_3\frac{\partial}{\partial\lambda_3}\right){\bf V}_{\rm eff}[\phi_0,\mu,\lambda_3]=0
 \label{rg equation}
 \end{align}
 Assuming that the effective action is dimensionless, and the effective potential therefore has dimension $3+\epsilon$, together with dimensional analysis leads to the following equation
  \begin{align}
 \left(\mu\frac{\partial}{\partial\mu}+\phi_0\frac{\partial}{\partial\phi_0}
 -3-\epsilon\right){\bf V}_{\rm eff}[\phi_0,\mu,\lambda_3]=0
 \label{dimensional analysis equation}
 \end{align}
Combining equations (\ref{rg equation}) and (\ref{dimensional analysis equation}) yields the the flow equation
 \begin{align}
 \left(\mu\frac{\partial}{\partial\mu}+\frac{(3+\epsilon)\gamma(\lambda_3) }{1-\gamma(\lambda_3)}
 +\frac{ \beta_3(\lambda_3)}{1-\gamma(\lambda_3)}\frac{\partial}{\partial\lambda_3}\right){\bf V}_{\rm eff}[\phi_0,\mu,\lambda_3]=0
\label{flow equation}
 \end{align}
 If we demand that the effective potential satisfies the flow equation
 (\ref{flow equation}) with boundary condition given by the subtraction scheme in formula (\ref{scheme}), we find the expression for the effective potential   
 \begin{align}
&{\bf V}_{\rm eff} =Nz(t)^{3+\epsilon}\left| \phi_0 
\right|^{3+\epsilon}
 \left[ \frac{ \Gamma[-\frac{3}{2}-\frac{\epsilon}{2}] }{(4\pi)^{\frac{3}{2}+\frac{\epsilon}{2}}}
 + {\rm sign}(\phi_0) \frac{\tilde\lambda_3(t)}{6}\right]
 \label{final potential}\\
 & t=\ln\frac{\mu}{|\phi_0| } \\
& \frac{d\tilde\lambda_3(t)}{dt}=-\frac{\beta_3(\tilde\lambda_3(t))}{1-\gamma(\tilde\lambda_3(t))} 
~,~~\tilde\lambda_3(0)=\lambda_3 \label{beta 1}
\\ &
z(t)=\exp\left( -\int_0^{t} ds\frac{\gamma(\tilde\lambda_3(s)) }{1-\gamma(\tilde\lambda_3(s))}\right)
 \label{z(t)} \end{align}
  In these equations, we denote the coupling by $\tilde\lambda_3(t)$ since it has a slightly different flow equation (\ref{beta 1}) from $\lambda_3$.   On the other hand, $\tilde\lambda_3$  has the same fixed points as $\lambda_3$ and its flow can terminate at, but it cannot cross zero, which is a fixed point. 
  
  If we take the large $N$ limit where $z(t)=1$ and $\tilde\lambda_3(t)=\lambda_3\frac{\mu^\epsilon}{|\phi_0|^\epsilon}$ the expression in equation (\ref{final potential}) reproduces the one that we found in that limit, equation (\ref{seff13d+epsilon}). 
  
  It is clear that, for any minimum of the potential in equation (\ref{final potential}) at a non-zero value of $\phi_0$ will  occur where $\phi_0$ and $\tilde\lambda_3$ (and therefore also $\lambda_3$) have opposite signs.   We shall, without loss of generality, assume that $\lambda_3$ and $\tilde\lambda_3$ are positive and that $\phi_0$ is negative.
To seek extrema of the effective potential in equation  (\ref{final potential}) $\phi_0$ we set its first derivative to zero.  This yields the equation
     \begin{align}
     &  \frac{1}{N}\frac{\partial}{\partial |\phi_0|} {\bf V}_{\rm eff} =0= \nonumber \\
    &  \frac{z^{3+\epsilon}|\phi_0|^{2+\epsilon}}{1-\gamma}
 \left[ (3+\epsilon)\frac{\Gamma[-\frac{3}{2}-\frac{\epsilon}{2}] }{(4\pi)^{\frac{3}{2}+\frac{\epsilon}{2}}}
-(3+ \epsilon ) \frac{\tilde\lambda_3}{6}- \frac{ \beta(\tilde\lambda_3)}{6}\right]
\label{equation for phi}
\end{align}
We note that, given that the quantum field theory can be renormalized so that its effective potential has the form
in equation (\ref{final potential}), no approximation has been made to obtain the equation (\ref{equation for phi}).   Of course,  
one solution of  equation (\ref{equation for phi}) is $\phi_0=0$.  
A second solution is found where the other factor vanishes, 
\begin{align}
 \frac{\Gamma[-\frac{3}{2}-\frac{\epsilon}{2}] }{(4\pi)^{\frac{3}{2}+\frac{\epsilon}{2}}}-
 \frac{\tilde\lambda_3(t)}{6}=
\frac{ \beta(\tilde\lambda_3)}{(3+\epsilon)6}
\label{solutionnnn}
\end{align}
 This is an equation for the coupling constant $\tilde\lambda_3$ with determines the value that the running coupling $\tilde\lambda_3(t)$ must flow to as a necessary (but not sufficient) condition that a nonzero condensate exists.
This equation tells us that the solution for the coupling depends very much on the beta function. In our case, we only know the beta function approximately at large $N$ and we know that it is small in that regime when $N$ is alare and $\epsilon$ is small.  In that regime, where the beta function is of order $\epsilon$ or $1/N$ and these are both somall, equation (\ref{solutionnnn}) is solved by 
\begin{align}
\tilde \lambda_3=\frac{1}{\pi}+\mathcal O(\epsilon,1/N)
\end{align}
We are therefore interested in renormalization group trajectories where   $\tilde\lambda_3(t)$ passes
near the point $1/\pi$.  

If we plug equation (\ref{solutionnnn})  back into the potential, we find that the on-shell energy of the non-zero solution  is
 \begin{align}
&\left. {\bf V}_{\rm eff}\right|_{\frac{\partial}{\partial |\phi_0|} {\bf V}_{\rm eff} } =Nz(t)^{3+\epsilon}\left| \phi_0 
\right|^{3+\epsilon}
 \left[  \frac{\beta(\tilde\lambda_3)}{(3+\epsilon)6}\right]
 \end{align}
 Again, this is an exact equation, without an approximation yet. 
This potential energy of the non-zero solution should be compared with  the energy of the $\phi_0=0$ solution which is $\left.  {\bf V}_{\rm eff}\right|_{\phi_0=0}=0$.  The nonzero solution is favoured when its energy is lower of the two and this is only when 
the beta function is negative in the vicinity of the solution
\begin{align}
\tilde\beta(\tilde\lambda_3\sim1/\pi)<0
\end{align}
with $\tilde\lambda_3$ evaluated at the solution of equation (\ref{solutionnnn}). Thus a stable nontrivial solution would seem to exist only when the beta function is negative  in the vicinity (with $\epsilon$ and $1/N$ accuracy) of the critical coupling $\tilde\lambda_3 \sim \frac{1}{\pi}$.  For stability of the solution, $\lambda_3\phi^3$ must be a marginally relevant operator.

Another way to see this is to examine the second derivative, which we find approximately
\begin{align}
& \left.  \frac{1}{N} \frac{\partial^2}{\partial|\phi_0|^2}{\bf V}_{\rm eff}  \right|_{\frac{\partial}{\partial |\phi_0|} {\bf V}_{\rm eff} =0}= 
    -   \frac{ |\phi_0|}{2}
  \beta_3(\tilde\lambda_3)
 +\mathcal O(\epsilon^2, \epsilon/N,1/N^2)
 \label{second derivative}
 \end{align}
 which also tells us that the locally stable nonzero solution must have the beta function negative.
 
  Our conclusion is, that the existence of solutions with a condensate, and overall stability of the model itself depend critically on the details of the beta function.  We have an approximation to the beta function in equation (\ref{beta function}) which is accurate when   $1/N$ is small.  We have been assuming that both 
  $\epsilon$ and $1/N$ are small and are of roughly the same magnitude, so that
$\epsilon N$ is of order one. Then, we can consider three cases which depend on the magnitude of $\epsilon N$.
 \begin{enumerate}
     \item{} When $\epsilon N > \frac{32}{\pi^2}$: There is an ultraviolet fixed point at $\lambda^{*UV}_3=0$ and  there is no infrared fixed point in the perturbative regime. As depicted in figure \ref{beta1}, the beta function is negative for all positive values of $\tilde\lambda_3$. In this regime the nonzero solution of equation (\ref{equation for phi}) is indeed a minimum. If we begin at $\phi_0\to\infty$ where $\tilde\lambda_3\to \tilde\lambda_3^{*UV}=0$ and lower $\phi_0$, $\tilde\lambda_3$ will increase until it satisfies equation  (\ref{equation for phi}). 
       \begin{figure}
\begin{center}
\includegraphics[scale=.5]{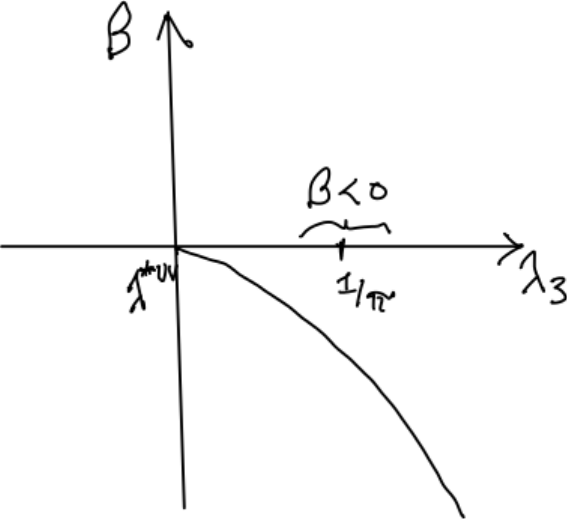}
 \caption{The beta function when $\epsilon N > \frac{32}{\pi^2}$.  The critical coupling occurs in 
 a region where the beta function is negative. The extremum of the effective potential found there is stable.   }
\label{beta1}
\end{center}
\end{figure}
     The coupling constant flow will then be cutoff in the infrared by the condensate $\phi_0$ itself which gaps the fermion spectrum. The mass of the scalar field is estimated from the curvature of the effective potential
     at this solution  $m_{\rm scalar}^2=12\pi \phi_0^2\beta(\lambda_3=\frac{1}{\pi})$ which is of order $\epsilon$ or  $1/N$ times the fermion mass squared\footnote{Both $\epsilon$ and $1/N$ can be arbitrarily small while holding $\epsilon N$ of order one and large enough to be in this regime.}  $\phi_0^2$, rather than order one as it could be when the curvature had its more natural magnitude $\sim N$. 
   \item{}When $1-\frac{8}{\pi^2}<\epsilon N<\frac{32}{\pi^2}$: There is an infrared fixed point at $\lambda^{*IR}_3=0$ and an ultraviolet fixed point at
     $\lambda_3^{*UV}=\sqrt{ \frac{1}{8}(1-\frac{\pi^2N\epsilon}{32})}$. As depicted in figure \ref{beta2}, the ultraviolet fixed point occurs at sufficiently weak coupling that the critical coupling is on the strong coupling side of that fixed point. The beta function is negative in the vicinity of the critical coupling.  The extremum of the effective potential found there is stable. 
      \begin{figure}
\begin{center}
\includegraphics[scale=.5]{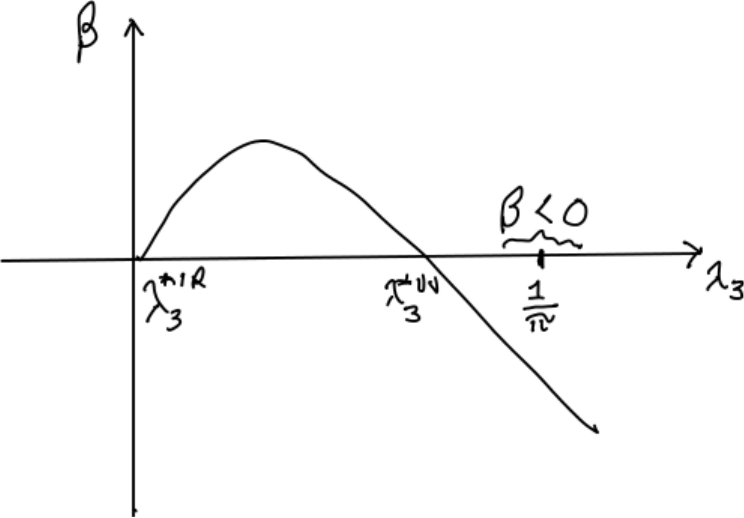}
 \caption{The beta function when $1-\frac{8}{\pi^2}<\epsilon N<\frac{32}{\pi^2}$ has an infrared fixed point and  an ultraviolet fixed point as depicted.  The ultraviolet fixed point occurs at sufficiently weak coupling that the critical coupling is on  the strong
 coupling branch.  In this case 
  the beta function is negative in a region near the critical coupling. The extremum of the effective potential found there is stable.  }
\label{beta2}
\end{center}
\end{figure}
\item{} When $\epsilon N < 1-\frac{8}{\pi^2}$:  The beta function has an infrared fixed point at zero and a nontrivial ultraviolet fixed point at a positive value of $\tilde\lambda_3$.  As depicted in figure (\ref{beta3}), in this case the critical coupling occurs between the two fixed points
in a region where the beta function is positive.   This means that the extremum of the potential that we find that is located near the critical coupling is a local maximum, rather than minimum.   The stable phase 
is one with runaway condensate $\phi_0\to \infty$ where $\tilde\lambda$ flows to the ultraviolet fixed point.  One can check that the effective potential is negative there.  There is also a metastable solution with $\phi_0=0$.
Thus we conclude that,
 when $\epsilon N<1-\frac{8}{\pi^2}$, which includes $\epsilon=0$, the case of exactly three dimensions, the theory is unstable.  
      \begin{figure}
\begin{center}
\includegraphics[scale=.5]{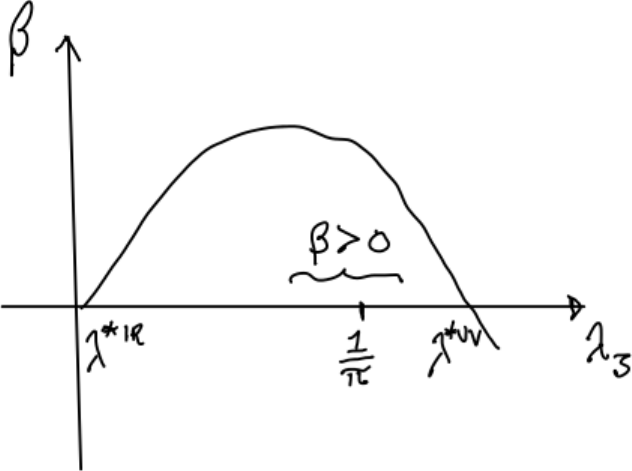}
 \caption{The beta function when $\epsilon N<1-\frac{8}{\pi^2} $.  The critical coupling occurs between
 the infrared and the ultraviolet fixed points, where the beta function is positive.  The theory is unstable in this case. }
\label{beta3}
\end{center}
\end{figure}
 \end{enumerate}

 In conclusion, we emphasize that, like the $\lambda_3\phi^3$ deformation of the critical Gross-Neveu model that we are discussing  is accessible by the large $N$ expansion, for any value of $\epsilon$ between zero and one, although we analyze it only in the large $N$ expansion where $\epsilon$ is of the order of or less than $1/N$.  However, when we restrict ourselves to $1/N$-perturbatively renormalizable quantum field theory, we obtain a stable well defined theory only for those values of $\epsilon $ and $1/N$ for which the deformation  $\lambda_3\phi^3$ is  marginally relevant.    Increasing $\epsilon$ helps in that it tends to make the operator more relevant.  The interactions of the theory -- nonlinear effects -- tend to make the operator less relevant.    If $\epsilon$ wins this competition, we find 
 a very nice structure.  The scalar field obtains a condensate, whose value becomes the fermion mass gap.  And a natural flatness of the potential give the fluctuations of the condensate a parametrically small mass $\sim\sqrt{\epsilon}$ times the fermion mass.  If $\epsilon$ loses this competition, the theory is unstable. Its solution has the fermion mass gap going to infinity.  There is a metastable solution with condensate zero and massless fermions which, however has a rather low barrier to its decay. 
 
Our analysis of the theory relies on the assumption that counter-terms can be added to the action so that the radiatively induced fermion mass, scalar field tadpole $\sim\phi$ and the quadratic term $\sim \phi^2$ are canceled at each order of $1/N$ perturbation theory.  This is indeed the case in our explicit computation up to order $1/N$    where we see the nontrivial fact that the scalar field tadpole is entirely of the form constant$\cdot\phi_0$ and not constant$\cdot|\phi_0|$ which would not qualify to be eliminated by a local counter-term which must be analytic in $\phi_0$. We expect that this good behaviour persists to higher orders, which would be needed for the model to be renormalizable.  It would be interesting to see whether one could develop an all-orders argument that this theory can be renormalized by analytic relevant and marginal counterterms only.

Finally, we have confirmed independently of our renormalization of the $1/N$ corrections to the effective potential, that in the same order of $1/N$ perturbation theory, the wave-function renormalization counterterms that we have used are indeed the appropriate ones to renormalize the Yukawa vertex and the counterterms renormalizing  the coupling $\lambda_3$ are indeed what is needed to remove the ultraviolet singularities from the scalar 3 point function. We do not include the details here as they are a very straightforward exercise to reproduce. 
 
  The authors acknowledge financial support of the Natural Sciences and Engineering Research Council of Canada.

 \end{document}